# Ultrafast decoupling of atomic sublattices in a charge-density-wave material

Jun Li[1], Junjie Li[1], Kai Sun[2], Lijun Wu[1], Haoyun Huang[1,3], Renkai Li[4], Jie Yang[4], Xiaozhe Shen[4], Xijie Wang[4], Huixia Luo[5], Robert J. Cava[5], Ian K. Robinson[1,6], Yimei Zhu[1,3], Weiguo Yin[1], Jing Tao[1,*]

[1]Condensed Matter Physics & Materials Science Department, Brookhaven National Laboratory, Upton, NY 11973, USA.

[2]Department of Physics, University of Michigan, Ann Arbor, MI 48109, USA.

[3]Department of Physics & Astronomy, Stony Brook University, Stony Brook, NY 11794, USA

[4]SLAC National Accelerator Laboratory, Menlo Park, CA 94025, USA.

[5]Department of Chemistry, Princeton University, Princeton, NJ 08544, USA.

[6]London Centre for Nanotechnology, University College, London WC1E 6BT, UK

Atomic rearrangements within crystals lie at the foundation of electron-phonon-coupled phenomena such as metal-insulator transition and superconductivity[1-5]. Advanced laser-pump-probe studies have recently focused on various charge-density-wave (CDW) materials to sharpen our understanding of the charge-lattice entanglement, where non-thermal melting of the CDW state is evident from the enhanced Bragg diffraction peak intensities—attributed to the dominance of the metal-atom dynamics over the nonmetal-anion one[6-10]. Here using ultrafast MeV electron diffraction[11-13] on the prototypical CDW material 1T-TaSeTe, we observe an unusual coexistence of systematically enhanced and suppressed Bragg peak intensities upon the CDW suppression, indicating a dominance of nonmetal-anion dynamics during photoexcitation. By tracking these atomic trajectories quantitatively through the ultrafast process, we identify a transient state that manifests itself as an unexpected decoupling of the Ta and Se/Te sublattices. These findings unambiguously unveil a new kind of laser manipulations of lattice order parameters, which has potentials in creating new quantum states and discerning hidden phases such as intra-unit-cell orders.

The crystal lattice response, seen as movements of all constituent atoms, to an ultrafast laser pulse on a material sample, allows understanding of the driving mechanisms and their underlying physical principles, particularly in CDW materials with quantum entanglement between charge and lattice[5-10]. Popular CDW materials consist of metal elements, e.g. V, Ta and La, and chalcogens as O, S, Se and Te. Many ultrafast investigations suggest the dominant role of metal elements in structural dynamics upon photoexcitation, supported by ultrafast scattering observations that depict the melting of the CDW phase with intensity enhancement of Bragg peaks[6-10]. This postulation on the metal-atom dynamics is based on the assumption by neglecting the response of the atomic sublattice of chalcogens. Indeed, the dynamics of chalcogen sublattice has never been clarified during the ultrafast processes in CDW materials. It led to a substantial deficiency in current knowledge of lattice dynamic, which is essential to exploit the laser manipulations of photoexcited materials.

We used MeV ultrafast electron diffraction (UED) techniques, which produce quality data that contain plenty of diffraction peaks at large momentum transfer in reciprocal space due to the large scattering cross-sections of electrons with momentum sensitivity, to investigate the structure dynamics of 1T-TaSeTe single-crystals, selected for the simple layer-stacking sequence of the 1T polymorph. We stabilized the crystal structure by replacing one Se with Te from its parent material, 1T-TaSe$_2$ crystal[14,15]. The 1T polymorph with $P\bar{3}m1$ symmetry, as illustrated in Fig. 1b, consists of quasi two-dimensional layers which are weakly bound to each other along the $c$ axis via van-der-Waals interaction. As far as we know, the Se and Te atoms are randomly distributed on the chalcogen sites in the hexagonal planes and covalently bounded with the Ta atoms. When the CDW phase and the CDW-induced periodic lattice distortion (PLD) are not taken into account, the electron structure factor of Bragg peak $\mathbf{g} = (hk0)$ can be written as

$$F_{\mathbf{g}} = \begin{cases} (f_{\text{Ta}} + 2f_{\text{Se/Te}}), & h-k = 3n \\ (f_{\text{Ta}} - f_{\text{Se/Te}}), & h-k \neq 3n \end{cases} \quad \text{with } n \text{ an integer,} \quad (1)$$

where $f_{\text{Ta}}$ and $f_{\text{Se/Te}} = (f_{\text{Se}} + f_{\text{Te}})/2$ (because of the same occupation probability of Se and Te on all chalcogen sites[15]) are the electron atomic scattering factors of Ta and chalcogen atoms, respectively. The Bragg peaks can be classified into two groups: ($h$-$k$) equals to $3n$ (labeled as Type-1) or not (Type-2), to which the Ta atoms have the same contribution but the Se/Te atoms' contributions are of opposite sign. When the incommensurate lattice modulations appear in the system[14,15], the small deviation of Ta atoms from the normal state (non-PLD; high symmetry) results in an intensity drop of *all* Bragg peaks without momentum-selection, as shown in Fig. 1b. By contrast, small displacements of Se/Te atoms give rise to a reduction of Type-1 peaks and an increase of Type-2 peaks (see Fig. 1c and Supplementary Fig. 1 for Bragg-peak-intensity variations as a function of Ta and Se/Te atomic displacements). The classification of Bragg peaks in Eq. (1) provides a new ground to test the contributions from different atomic motions during dynamic processes in the CDW materials with metal and chalcogen elements in hexagonal lattices. Particularly, it confirms the dominant role of the Ta-sublattice dynamics in photoexcited TaS$_2$[6,9], where all the Bragg peaks intensities were measured to behave similarly through the melting and recovery process of the CDW/PLD state, i.e., the motif in Fig. 1b. On the other hand, it should clearly indicate the primary role of the chalcogen sublattice during the dynamics if the motif in Fig. 1c is observed, as the results obtained in this study.

A typical MeV-UED pattern of 1T-TaSeTe taken along the *c* axis at the temperature of 26 K is shown in Fig. 2a. The pattern contains both Bragg peaks and PLD satellite reflections, illustrating a "triple-**q**" PLD state. The inset of Fig. 2a shows in-plane components of the three PLD wavevectors which are linear combinations of the reciprocal lattice vectors (***a****$^*$, **b**$^*$ and ***c***$^*$) that $\mathbf{q}_1 = q \cdot \mathbf{a}^*$, $\mathbf{q}_2 = -q \cdot \mathbf{b}^*$ and $\mathbf{q}_3 = q \cdot (-\mathbf{a}^* + \mathbf{b}^*)$, with q ~ 0.288 at 26 K. This type of PLD satellites is also found in 2H-TaSe$_2$ at low temperatures (90-122 K) and 1T-TaS$_2$ at high temperatures (350-550 K)[8,16]. The time-resolved normalized integrated intensities of the Bragg peaks from the UED data are shown in Fig. 2b and 2c. The measurements are sorted into Type-1 and Type-2 with the same momentum distribution as described in Eq. (1) and illustrated in the left half of Fig. 2a. This behavior is distinct from the previous reports in the family compounds such as 1T-TaS$_2$, 2H-TaSe$_2$, LaTe$_3$ and 4H$_b$-TaSe$_2$[6-10], but identical to the motif in Fig. 1c. Note that the intensity trend of Type-1 and Type-2 before $t$ ~ 400 fs is precisely opposite to the motif shown in Fig.1c for motions of Se/Te atoms to form a PLD. The opposite direction tells us that the atoms are set moving away from the PLD state towards the non-PLD state, i.e., through a melting process of PLD upon photoexcitation. Since only the motion of Se/Te atoms can cause this pattern of intensity variations, our UED results indicate that the motion of Se/Te atoms plays a dominant role at the early stage of lattice evolution, contrary to the widely reported motions of metal atoms[5-10,17].

We have been able to reconstruct the in-plane components of the evolution of the 1T-TaSeTe crystal structures by fitting a model of the atomic coordinates (see Supplementary Information for details). Firstly, the UED patterns display quasi-kinematic characteristics, owing to the extremely high electron accelerating voltage (3.5 MeV, leading to less multiple scattering[13,18]) and the rippling of the thin sample foil within the large electron illuminating area (~300 μm as the diameter, leading to a similar effect as precession electron diffraction[19]), Indeed, dynamic electron simulations of this material and the comparison with the experimental results strongly suggest that, in this case, the dynamical scattering effects only add minor perturbation to the total scattering intensities of the kinematic simulations (Supplementary Fig. 2). Basically, the PLD structural model in each atomic layer is constructed assuming a sinusoidal displacement wave along each of the three symmetry equivalent wave vectors $\mathbf{q}_1$, $\mathbf{q}_2$ and $\mathbf{q}_3$[20]. We further identify a predominantly longitudinal mode in PLD atomic displacements in 1T-TaSeTe by the features illustrated in the inset of Fig. 2a that only four of six PLD reflections appear around ($1\bar{2}0$) peak (Supplementary Fig. 3). Considering the lack of spatial resolution along the ***c**** direction of the UED patterns, the number of parameters can be reduced to three: the amplitude $u_{Ta}$ of the Ta modulation wave, the amplitude $u_{Se/Te}$ and phase $\varphi_{Se/Te}$ of the Se/Te modulation wave ($\varphi_{Se/Te}$ is opposite in two adjacent Se/Te layers separated by Ta). Note that the wave numbers of $\mathbf{q}_1$, $\mathbf{q}_2$ and $\mathbf{q}_3$ are incommensurate so the unit cell of the PLD is more complicated than a 3×3 PLD unit-cell constructed by density-functional-theory (DFT) methods[21]. To accommodate these PLD features fully into the diffraction model, it was necessary to include modulated atomic scattering factors[20] and anisotropic Debye-Waller factors[22]. Least-squares refinement[23] was employed using the incommensurate crystallography method, described as a projection from higher dimensional space[20].

As a result, Fig. 3a shows a good agreement in a comparison of the peak intensities, for both Bragg peaks and PLD satellites, between the experimental data and the simulated pattern (inset) using the extracted parameters for delay time $\Delta t$ = -2 ps. The displacements of Ta, Se/Te atoms with respect to the non-PLD state and the relative phase between the two sets of atoms are plotted in Fig. 3b as a function of time delay. Before the photoexcitation, Ta and Se/Te atoms are found at their PLD positions. As the system is optically excited, both the Ta and Se/Te atoms move from the starting PLD positions toward the undisplaced (non-PLD) positions in the first 400 fs and then start to relax back to partially recovered positions after $t$ ~ 3 ps. The meta-stable state can last up to 1 ns based on the experimental observations at longer delays. Meanwhile, the phase parameter $\varphi_{Se/Te}$ remains nearly unchanged during this process.

This lattice excitation and relation behavior unveil a short-time transient state. Fig. 3b shows that, at $t$ ~ 400 fs, Ta atoms have moved back from the original PLD distortion by only ~ 6.3% while Se/Te atoms have moved ~ 41%, nearly the half way from the PLD state to the high-symmetry non-PLD state. Later on at ~ 3 ps, after the transient has passed, the distortions are back roughly in proportion to form a reduced-amplitude version of the PLD state (~ 90% for both the elemental sublattices). So, while the laser excitation eventually couples to the PLD distortion, reducing its magnitude, it does this via an intermediated state with different distortions, mostly on the Se/Te site.

The formation of the symmetry breaking PLD state in this material is considered to arise from softening of a single phonon mode in an acoustic branch upon cooling[24-26]. If this photoexcitation process were driven harmonically with the single phonon mode, the ratio of the relative distortions of the Ta and Se/Te sublattices would be exactly in proportion. However, experimental finding Fig. 4a shows a drastically different distortion pattern. The fact that the initial deformation only couples later to the mode associated with the PLD indicates that, in the transient state, either multiple phonon modes are responsible to the lattice deformations or there is a pronounced anharmonicity in the single mode response to a moderate pumping fluence of 3.5 mJ/cm$^2$. Specially, the existence of the transient state is unlikely to be rooted in multiple modes because different time scales would be observed, and the atomic trajectories would probably not be straight lines, both are contrary to the experimental observations. On the other hand, the free energy of an elastic wave is proportional to square of the wave amplitude, which is a harmonic term. Anharmonic effects normally comes from higher order terms in the free energy, usually from the quartic order. However, the "triple-$q$" state of the PLD in this material gives rise to a special anharmonic term which is cubic in the wave amplitude $\phi_{\mathbf{q_1}}\phi_{\mathbf{q_2}}\phi_{\mathbf{q_3}}$[27]. This cubic term breaks the symmetry between $\phi$ and $-\phi$, making the energy landscape asymmetric as illustrated in Fig. 4d. Since a cubic term in energy is much stronger than quartic ones for small deformations, anharmonic responses can be substantial in this material, even under moderate pumping fluence.

The appearance of a transient state at $t$ ~ 400 fs illustrates a trend of the decoupling of Ta and Se/Te sublattices in terms of their symmetry-breaking status, seen in Fig. 4d, in which the Ta sublattice remains at low symmetry (PLD state in Fig. 4c) while the Se/Te sublattice evolves toward high symmetry (normal state in Fig. 4b). Transient "hidden phases"[28-30], are known to involve the decoupling of sublattices of different orders (charge, spin and lattice). Our findings indicate that 1T-TaSeTe, prompted by laser pulses, shows a breaking of the lattice order into subgroups (atomic sublattices) as an intra-unit-cell order in the ultrafast time regime. Neither the evolution of the Ta sublattice nor that of Se/Te sublattice can represent the complete structural dynamics of this material. Consequently, order parameters of the lattice distortions would be associated with the amplitudes and phases of the sublattices, in which key characteristics of structure and phonon information are encoded. Moreover, the most drastic structural anomaly takes place when the thermal effect is minor (Fig. 4a), which indicates that the decoupling of the sublattices is manipulated by the photoexcitation as a fundamental control of order parameters with potentials of generating novel quantum states.

## Methods

**Sample preparation and experimental setup.** The 1T-TaSeTe single crystals were grown by chemical vapor transport (CVT) with iodine as a transport agent[18]. The single crystalline foils with an average thickness around 50 nm and diameter around 800 μm were mechanically exfoliated from the bulk using adhesive tape, and transferred to the nickel TEM meshes with the aid of acetone. The experiments reported here were performed on the 4.0-MeV-UED setup at the Stanford Linear Accelerator Laboratory which can achieve 100 fs temporal resolution. The principle of MeV-UED operation and technical details can be found in Ref. 8. Using this experimental setup, the free-standing foils were uniformly excited by a 60-fs [full width at half maximum (FWHM)] laser pulse with a photon energy of 1.55 eV (center wavelength of 800 nm) and a fluence of 3.5 mJ/cm$^2$. The optical pulses at a repetition rate of 120 Hz were focused down to 1.5 mm on the sample with uniform thickness to trigger electron and lattice dynamics. The laser pump pulse and probing electron beam are colinear in the UED instrument. The excited states within an illuminated area of diameter 400 μm were probed by well-synchronized 3.5 MeV electron pulses containing ~ 10$^6$ electrons, ensuring a large number of accessible reflections were recorded in each diffraction pattern. The sample temperature of 26 K was achieved using a conducting sample holder cooled by liquid helium. The selected area electron diffraction patterns included in the Supplementary Figure 2 were obtained with a JEOL ARM 200F microscope operating at 200 kV @ BNL.

**Crystal structure refinement using UED patterns of 1T-TaSeTe.** See supplementary information for the methods.

**Data availability.** The data that support the plots within this paper and other findings of this study are available from the corresponding author upon request.

## Acknowledgements

We thank Tatiana Konstantinova and Dr. Mark Dean for the discussion of this work. Research was sponsored by DOE-BES Early Career Award Program and by DOE-BES under Contract DE-SC0012704. The work at University of Michigan was supported by NSF-EFMA-1741618. The UED experiment was performed at SLAC MeV-UED, which is supported in part by the DOE BES SUF Division Accelerator & Detector R&D program, the LCLS Facility, and SLAC under contract

Nos. DE-AC02-05-CH11231 and DE-AC02-76SF00515. The work at Princeton was supported by Department of Energy, Division of Basic Energy Sciences, Grant DE-FG02-98ER45706.

**Author Contributions**

J.T. and Jun L. designed the project and were the primary interpreters of the resulting data. R.J.C. and H.L. synthesized the single-crystal material. J.T., Junjie L., R.L., J.Y., X.S. and X.W. collected MeV UED data. Jun L., J.T., Junjie L., W-G.Y., K.S. and Y.Z. analyzed UED data. W-G.Y. performed DFT calculations on the CDW state. All authors contributed to the writing of the paper.

Correspondence and requests for materials should be addressed to jtao@bnl.gov.

# Figures

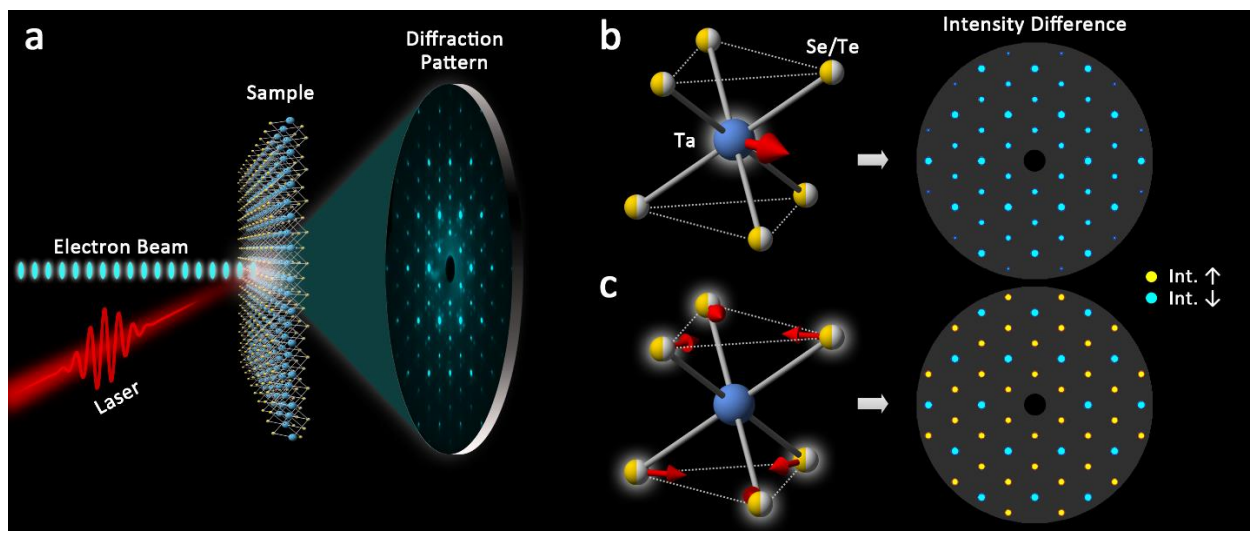

**Fig. 1 | Experimental setup and intensity change of Bragg peaks induced by atomic motions upon photoexcitation. a,** Schematic of the pump-probe experiment. The photo-induced electron pulses (blue) and the laser pulses (red) interact with the two-dimensional 1T-TaSeTe sample with tunable delay times. The laser pump pulse and probing electron beam are colinear in the actual UED instrument. **b,** The periodic in-plane displacement of Ta atoms (indicated by the red arrow) from the normal state without lattice distortions leads to the reduction of intensity of all Bragg peaks (blue spots). **c,** In the scenario with in-plane Se/Te displacements (red arrows), the intensities of some of the Bragg peak intensities become enhanced (yellow spots; indexed as (100), (200), *etc.*) while the others become weaker (blue spots; indexed as (110), (300), *etc*).

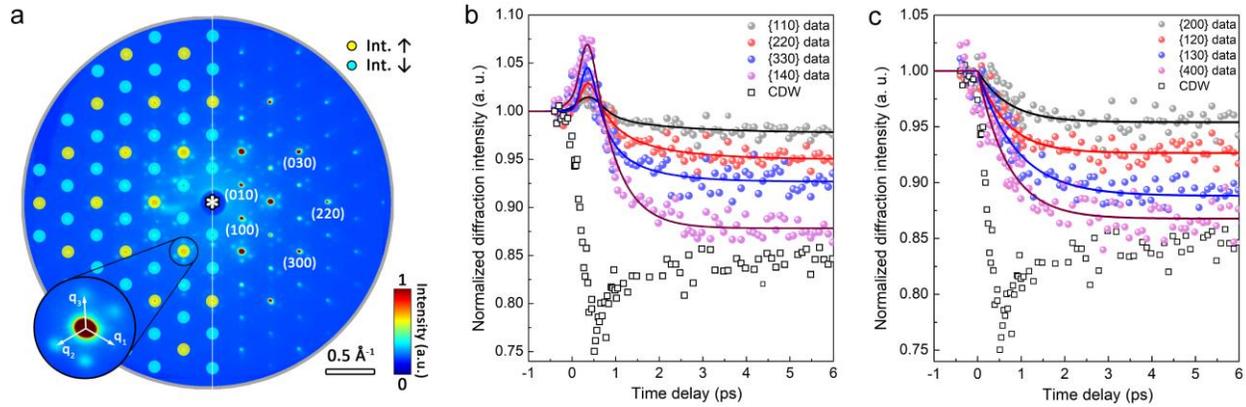

**Fig. 2 | Two types of time-dependent behaviors of the Bragg peak intensities after photo-excitation. a,** A typical UED pattern of single-crystalline 1T-TaSeTe obtained at 26 K. Five primitive Bragg peaks are indexed. Each Bragg peak is surrounded by six first-order incommensurate PLD satellites with different intensities. The adjacent area around the $(1\bar{2}0)$ peak is enlarged and shown in the inset. The three PLD wave vectors are indicated by arrows. The Bragg peaks whose intensities increase after photo-excitation are masked by yellow disks (Type-1) on the left half pattern, and the ones with decreased intensities are masked by blue disks (Type-2). **b,** The normalized intensity change of four Type-1 Bragg peaks as a function of pump-probe delay time. To improve the signal-noise-ratio, the intensity of each peak (colored circles) shown here is averaged from all symmetry equivalent peaks. The highest intensity is reached at ~ 400 fs for all the Type-1 Bragg peaks, and then the intensities decrease to reach a meta-stable value after ~ 3 ps. **c,** The normalized intensity change of four Type-2 Bragg peaks as a function of delay time. By contrast to Type-1, the intensities fall off right after the photo-excitation. Solid lines in **b** and **c** are eye guides for the measurements. For comparison, black boxes in **b** and **c** are typical intensity measurements from CDW/PLD reflection $(120_3)$ (see Fig. 3 for the notation of PLD-reflection indexing).

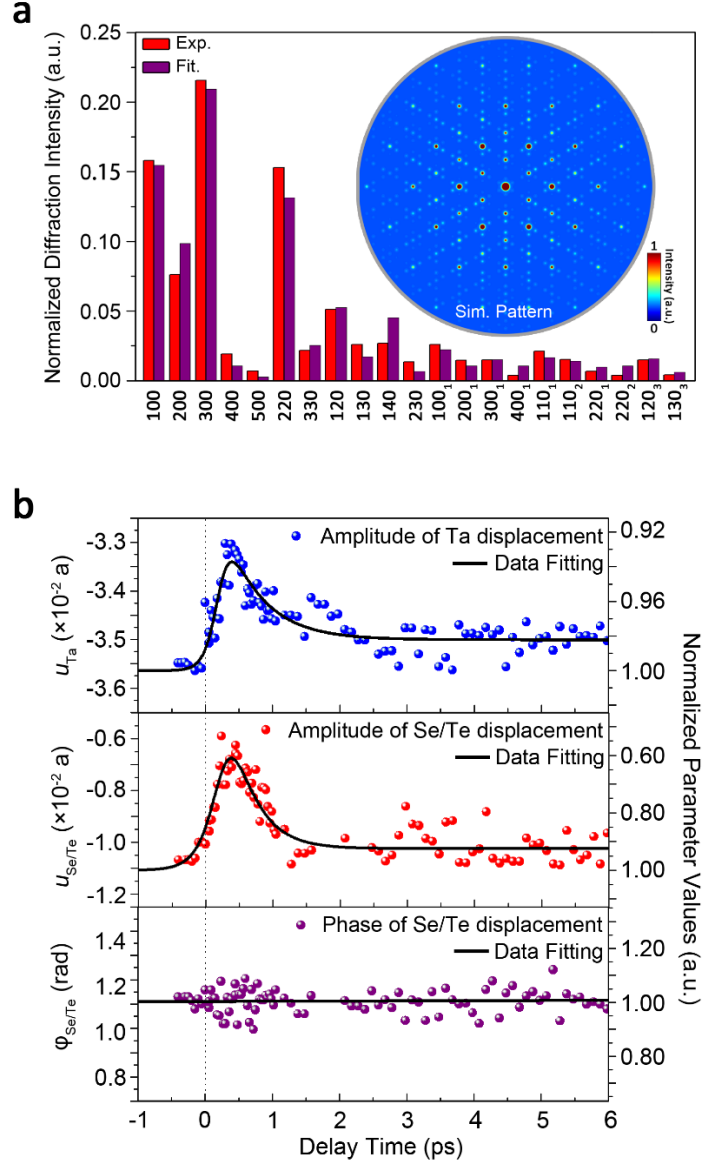

**Fig. 3 | Temporal Evolution of periodic lattice distortion (PLD) parameters extracted from experimental data. a,** A comparison between the experimental and kinematically simulated diffraction intensities of Bragg peaks and PLD satellites (indexed with subscripts). The experimental data was acquired before the photoexcitation (delay time $\Delta t$ = -2 ps) and the simulated fitting is based on the incommensurately modulated structure model described in the Supplementary Materials. The subscripts 1, 2 and 3 correspond to indexes ($hkl$100), ($hkl$010) and ($hkl$0$\bar{1}$0), respectively. The calculated diffraction pattern is shown in the inset, showing good agreement with the experimental pattern (Fig. 2a). **b,** Temporal evolution of PLD parameters $u_{Ta}$, $u_{Se/Te}$ and $\varphi_{Se/Te}$, respectively. The solid lines fit the data as eye-guides. $u_{Ta}$ is the amplitude of Ta modulation wave, $u_{Se/Te}$ is the amplitude of Se/Te modulation wave and $\varphi_{Se/Te}$ is the relative phase between the Se/Te and Ta sublattices.

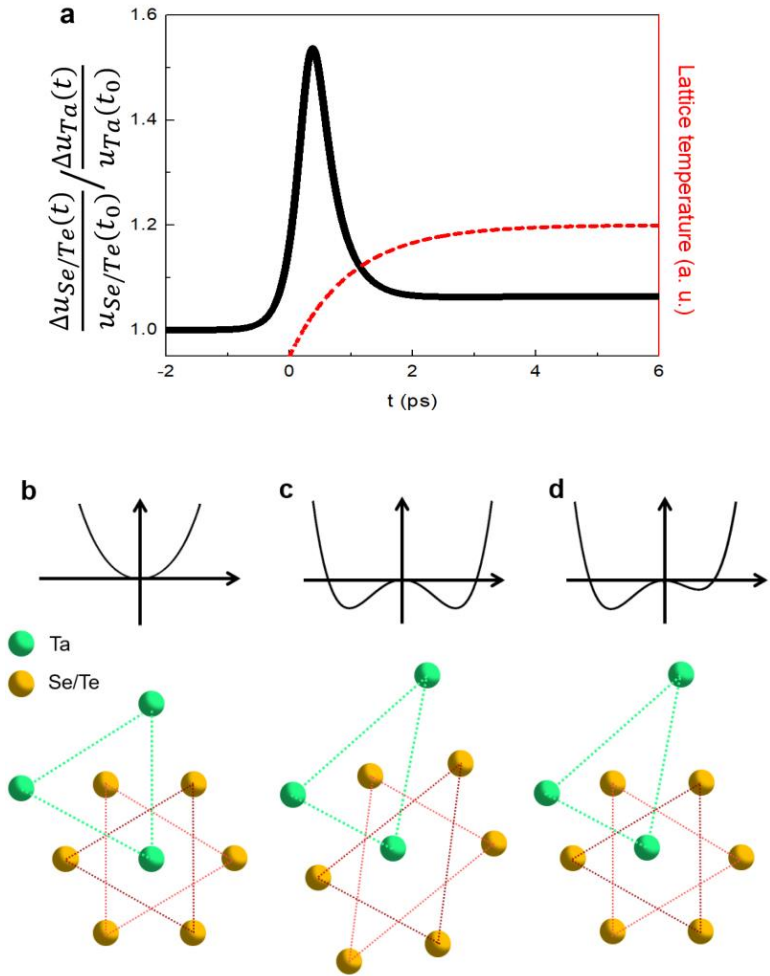

**Fig. 4 | Schematic representation of decoupling of the broken symmetry of sublattices in 1T-TaSeTe. a,** The black curve is the ratio of the relative displacements of Se/Te and Ta atoms $\frac{u_{Se/Te}(t_0)-u_{Se/Te}(t)}{u_{Se/Te}(t_0)}/\frac{u_{Ta}(t_0)-u_{Ta}(t)}{u_{Ta}(t_0)}$ as a function of time, calculated from Fig. 3**b**. The red dashed line is a fitting curve of the diffuse scattering measurements in background intensity of UED patterns (see Supplementary Fig. 9) and indicates a gradual increasing of the lattice temperature. For **b**, **c** and **d**, top row: energy landscape; bottom row: atomic arrangement. **b,** In the normal state without PLD, both the Ta and Se/Te sublattice are at high-symmetry positions. **c,** In the PLD state, both the Ta and Se/Te sublattice are at low-symmetry positions. **d,** The Ta sublattice is at the low-symmetry positions, while the arrangement of Se/Te sublattice goes to high-symmetry. Note that the higher energy minimum in the energy landscape is indeed a saddle point in 3D space.

# Supplementary Information:

# Ultrafast decoupling of atomic sublattices in a charge-density-wave material

Jun Li[1], Junjie Li[1], Kai Sun[2], Lijun Wu[1], Haoyun Huang[1,3], Renkai Li[4], Jie Yang[4], Xiaozhe Shen[4], Xijie Wang[4], Huixia Luo[5], Robert J. Cava[5], Ian K. Robinson[1,6], Yimei Zhu[1,3], Weiguo Yin[1], Jing Tao[1,*]

[1]Condensed Matter Physics & Materials Science Department, Brookhaven National Laboratory,

Upton, NY 11973, USA.

[2]Department of Physics, University of Michigan, Ann Arbor, MI 48109, USA.

[3]Department of Physics & Astronomy, Stony Brook University, Stony Brook, NY 11794, USA

[4]SLAC National Accelerator Laboratory, Menlo Park, CA 94025, USA.

[5]Department of Chemistry, Princeton University, Princeton, NJ 08544, USA.

[6]London Centre for Nanotechnology, University College, London WC1E 6BT, UK

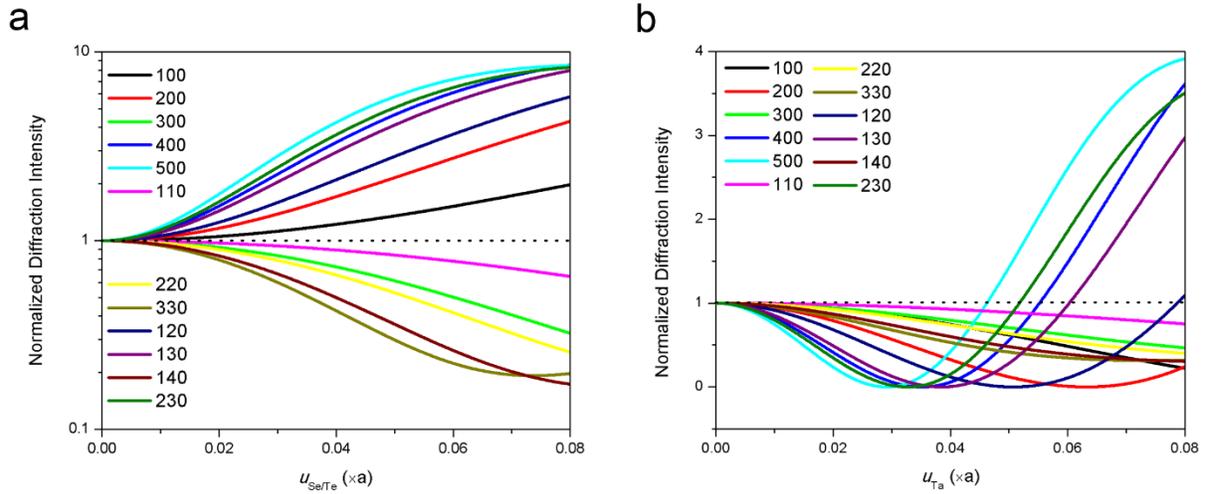

**Supplementary Figure 1 | Normalized diffraction intensities of Bragg peaks as a function of structural parameters.** Parameters $u_{Ta}$ and $u_{Se/Te}$ are the amplitudes of Ta and Se/Te displacement waves, respectively. These values can be negative if the phase terms of the waves are considered: $-u = u \cdot e^{i\pi}$. Calculated intensities in **a** and **b** based on the equation (7) in Section b), in which the intensities of Bragg peaks are independent on the phase parameter $\varphi_{Se/Te}$. **a**, The parameter $u_{Ta}$ is fixed to be 0 when $u_{Se/Te}$ varies from 0 to 0.08 of lattice parameter a, two types of Bragg peaks can be identified from the intensity plots. **b**, The parameter is $u_{Se/Te}$ fixed to be 0 when $u_{Ta}$ varies from 0 to 0.08 (×a), all the intensities decrease at $u_{Ta}$ small (< 0.03). It is clear that our measurements are qualitatively consistent with the scenario described in **a**, two distinct Bragg peaks can be identified, indicating the dominance of the movement of Se/Te sublattice during photoexcitation process.

**Crystal structure refinement using UED patterns of 1T-TaSeTe**

We herein show the methods for four key steps in order to extract atomic coordinates of Ta and Se/Te atoms from the UED patterns at each time delay. The four steps are summarized below, followed by elaboration of each step.

a) Symmetry consideration of the atomic displacement in 1T-TaSeTe. This step helps to reduce the fitting parameters in the later procedures.

b) Structure scattering factors using incommensurate superlattice modulations. Since the wave number of the superlattice reflections (SLRs) was measured to be at incommensurate values ($q \sim$ 0.29), electron diffraction simulations (with additional dimensions for incommensuration) offer accurate wave number of the SLRs, which assists the comparison between the simulations and experimental results.

c) Peak intensity measurements using two-dimensional rotated Gaussian functions to precisely determine the peak integrated intensity, peak position and peak width from the experimental UED patterns.

d) Electron diffraction simulations and structure refinement. We performed both dynamic and kinematic electron diffraction simulations. After compared with experimental data, we use kinematic simulations for the crystal refinement because the dynamic simulations show only minor perturbations to the kinematic simulations in this case.

**a) Symmetry of atomic displacements in 1T-TaSeTe.** The atomic displacements involved in the CDW transition are closely related to the symmetry properties of the normal vibrational modes. The group-theoretical methods have been developed to determine the symmetry properties of the normal modes of vibration and the corresponding forms of the eigenvectors[31]. In the case of 1T-TaSeTe, the structural modulation wave vectors $\mathbf{q}_1$, $\mathbf{q}_2$ and $\mathbf{q}_3$ are along the $\Gamma$-$\Sigma$-$M$ direction in the reciprocal lattice, and the crystallographic point group of the undistorted structure is $D_{3d}$[14]. The maximal point group whose symmetry operations leave a $\mathbf{q}$ vector invariant is $C_s$, consisting of the identity operation $E$ and a mirror plane $\sigma$. As shown in the character table for the $C_s$ group (Supplementary Table 1), there are two one-dimensional representations labeled $A'$ and $A''$.

**Supplementary Table 1 | Character table for group $C_s$.** The maximal group which leaves the **q** vector invariant is $C_s$. Two elements $E$ and $\sigma$ are contained in this group, indicating that there are two one-dimensional irreducible representations labeled A' and A". Thus, two kinds of atomic displacement symmetry can be identified in 1T-TaSeTe.

| $C_s$ | $E$ | $\sigma$ |
|---|---|---|
| A' | +1 | +1 |
| A" | +1 | -1 |

**Supplementary Table 2 | Transformation properties of atoms and Cartesian coordinates under the operation of $C_s$ and inversion.** This table is generated from the atomic configuration shown in Supplementary Fig. 4. With the aid of results summarized here, the $\Gamma(R)$ matrices can be derived, and the number of independent eigenvector components can be significantly reduced. For either A' or A" symmetry, only three independent quantities are sufficient to describe the in-plane atomic displacements.

|   | 1 | 2 | 3 | $x$ | $y$ | $z$ | $\chi^{\Gamma}(R)$ |
|---|---|---|---|---|---|---|---|
| $E$ | 1 | 2 | 3 | $x$ | $y$ | $z$ | 9 |
| $\sigma$ | 1 | 2 | 3 | $x$ | $-y$ | $z$ | 3 |
| $i$ | 1 | 3 | 2 | $-x$ | $-y$ | $-z$ |   |

By adopting the point-atom approximation and assuming that the modulation can be regarded as a sinusoidal plane wave, the atomic displacement for the $k$th atom in the $l$th unit cell is giving by[32,33]

$$u_{kl} = \text{Im} \sum_{\mathbf{q}} \frac{1}{2} \{u_k(\mathbf{q}) \exp[i\mathbf{q} \cdot (\mathbf{n}_l + \mathbf{r}_k)] - u_k^*(\mathbf{q}) \exp[-i\mathbf{q} \cdot (\mathbf{n}_l + \mathbf{r}_k)]\}$$

(1)

where $u_k(\mathbf{q})$ is the displacement vector for a $k$th atom in a normal mode of wave vector $\mathbf{q}$, $\mathbf{n}_l$ is the position vector of the origin of the $l$th unit cell and $\mathbf{r}_k$ is the position of the $k$th atom relative to the origin of the cell, and Im denotes the imaginary part. In the undistorted 1T structure, there are 3 atomic sites in each unit cell and $k = 1, 2, 3$ correspond to Ta, (Se/Te)$_1$ and (Se/Te)$_2$, respectively (see Supplementary Fig. 4). Therefore the general displacement eigenvector has 9 components which can be written as $u^j(\mathbf{q}) = (u_1(\mathbf{q}); u_2(\mathbf{q}); u_3(\mathbf{q}))^j = (x_1, y_1, z_1; x_2, y_2, z_2; x_3, y_3, z_3)^j$, where all the components can be complex and the superscript $j = 1, 2$ denotes the symmetry character A' and A" respectively. The relationship among these components can be revealed by considering the transformation properties of operators $R = E$ or $\sigma$ on the displacement eigenvector: $\Gamma(R)\cdot \boldsymbol{u}^j(\mathbf{q}) = \chi^j(R)\boldsymbol{u}^j(\mathbf{q})$, $\chi^j(R)$ is the character of $R$ in the $j$th representation. The transformations of the three atoms under each operation are summarized in Supplementary Table 2. When $j = 1$, we have $\chi^1(E) = \chi^1(\sigma) = 1$, and $\Gamma(E)\cdot\boldsymbol{u}^1(\mathbf{q}) = \Gamma(\sigma)\cdot\boldsymbol{u}^1(\mathbf{q}) = \boldsymbol{u}^1(\mathbf{q})$. By using Supplementary Table 2, we can write $(x_1, y_1, z_1; x_2, y_2, z_2; x_3, y_3, z_3)^1 = (x_1, -y_1, z_1; x_2, -y_2, z_2; x_3, -y_3, z_3)^1$, yielding $y_1 = y_2 = y_3 = 0$. In addition, as described in Ref. 31, the eigenvectors obey $\boldsymbol{u}_k^*(\mathbf{q}) = \boldsymbol{u}_k(-\mathbf{q})$ and $\boldsymbol{u}_k(-\mathbf{q}) = -\Gamma(i)\boldsymbol{u}_k(\mathbf{q})$. With the help of Supplementary Table 2, these equations give $(x_1^*, 0, z_1^*; x_2^*, 0, z_2^*; x_3^*, 0, z_3^*)^1 = (x_1, 0, z_1; x_3, 0, z_3; x_2, 0, z_2)^1$, yielding $x_1$ and $z_1$ are real numbers and $x_2^* = x_3$, $z_2^* = z_3$. As a result, the displacement eigenvector with $j = 1$ can be written as $(x_1, 0, z_1; x_2, 0, z_2; x_2^*, 0, z_2^*)^1$ which includes 6 independent components. Similarly, the eigenvector with $j = 2$ can be simplified to $(0, y_1, 0; 0, y_2, 0; 0, y_2^*, 0)^2$ where $y_1$ is a real number. Because only the diffraction peaks within the zero order Laue zone are contained in our diffraction patterns, the atomic displacements along the $\boldsymbol{c}$ axis do not contribute to the intensity changes. Thus only three independent components are needed to describe the in-plane lattice distortions of either $A'$ or $A''$ symmetry. Comparing the simulated diffraction patterns shown in Supplementary Fig. 3 with the experimental pattern (Fig. 2a), the intensities of PLD peaks indicate that the in-plane atomic displacements in 1T-TaSeTe are longitudinal and have $A'$ symmetry. As we used in the main text, the independent components can be written as $x_1 \equiv u_{Ta}$, $x_2 \equiv u_{Se/Te}\cdot\exp(i\varphi_{Se/Te})$.

The number of times that each irreducible representation is contained in the reducible representation $\Gamma$ can be expressed as $a_j = \frac{1}{h}\sum_R g(R)\chi^j(R)\chi^\Gamma(R)$, where $h = 2$ is the order of group $C_s$, $g(R) = 1$ is the number of operations in the class, and $\chi^\Gamma(R)$ giving in Supplementary Table 2 is the character of the $R$th operation in the reducible representation. Thus the reducible representation $\Gamma$ can be decomposed as $\Gamma = 6 A' + 3A''$, indicating that among the 9 phonon branches, six will have $A'$ symmetry and three will have $A''$ symmetry.

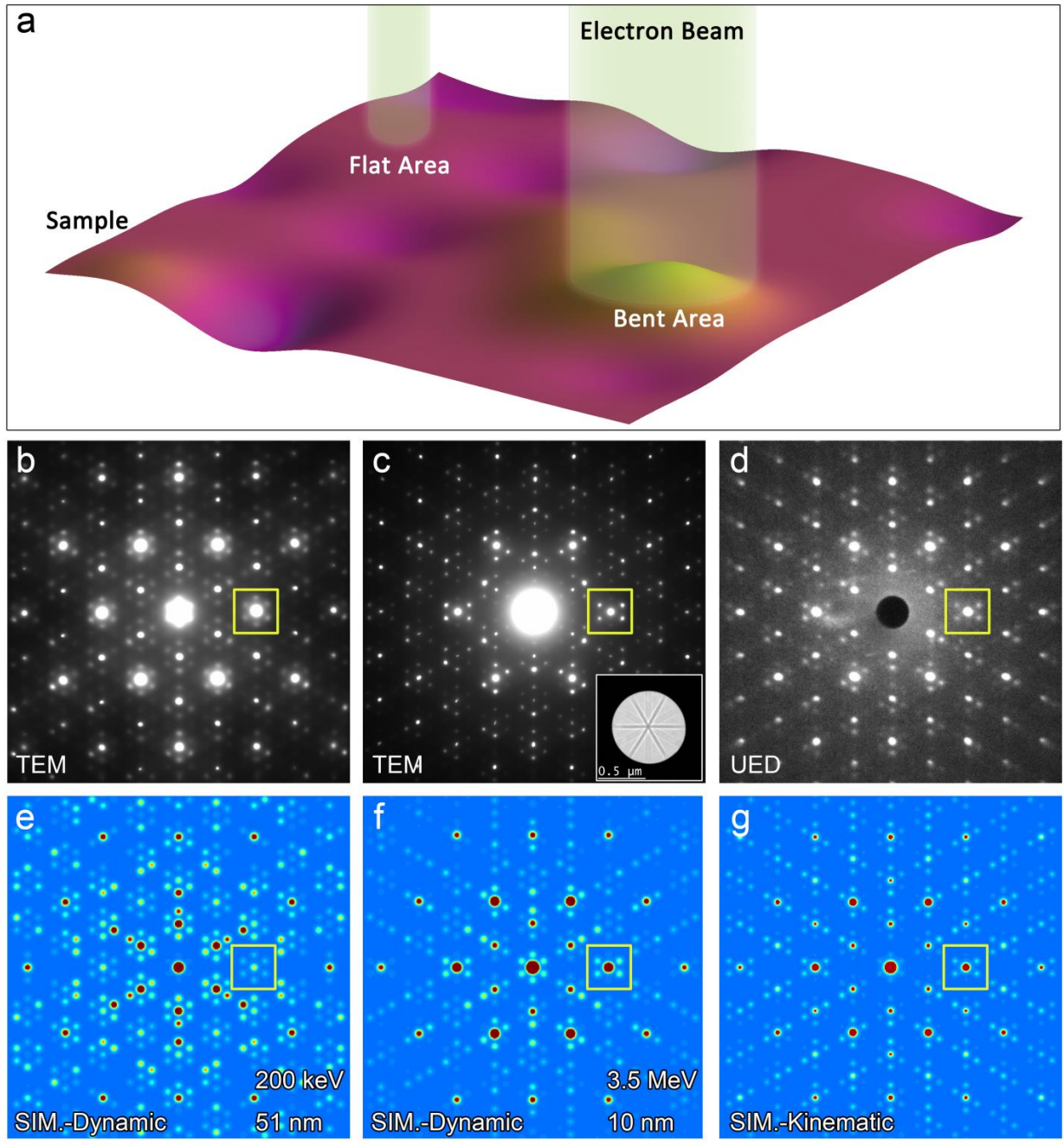

**Supplementary Figure 2 | Comparison between kinematic and dynamic diffraction. a**, A schematic drawing to show the UED sample morphology and electron diffraction acquisition. **b**, TEM diffraction pattern acquired from a flat area of the TaSeTe sample demonstrated in **a**, with the area diameter ~ 100 nm. Six CDW satellites around the (110) spot (within the yellow box) are at about the same intensity due to multiple electron scattering (dynamic diffraction effects). **c**, A TEM diffraction pattern acquired from a rippled region of the same sample demonstrated in **a**, with the area diameter ~ 700 nm. The inset shows the corresponding bright-field electron diffraction contrast image of the region with a symmetric bending contour (i.e., the diffraction obtained from the entire rounded valley or protuberance, the circle indicates the position of

selective area aperture). Four of the six satellites are much brighter than the other two. Because the sample area bends symmetrically so that the obtained electron diffraction has a similar effect as the precession diffraction technique, i.e., the ensemble of intensities in the pattern behaves in a "kinematick-like" fashion. **d**, A typical UED pattern which shows similar intensity distribution as presented in **c**, especially the CDW satellites around the (110) spot. **e**, Simulated electron diffraction pattern based on the Bloch wave theory with the thickness of 51 nm (accelerating voltage = 200 kV) that shows a good fitting to the experimental pattern (obtained at 200 kV) in **a**. The intensities of the six satellite spots are almost equal, indicating that the dynamic diffraction effect can result in this intensity feature. **f,** Simulated electron diffraction pattern based on the Bloch wave theory with the thickness of 10 nm (accelerating voltage = 3.5 MeV). **g**, Simulated electron diffraction pattern based on the kinematic theory. A comparison between the dynamic simulation in **f** and the kinematic simulation in **g**, using characteristics in the yellow box, shows similar intensity distributions in the patterns. Indeed, the dynamic scattering in our case only adds minor perturbation in intensity derived from the kinematic simulation, which qualitatively matches well with the experimental patterns shown in **c** and **d**.

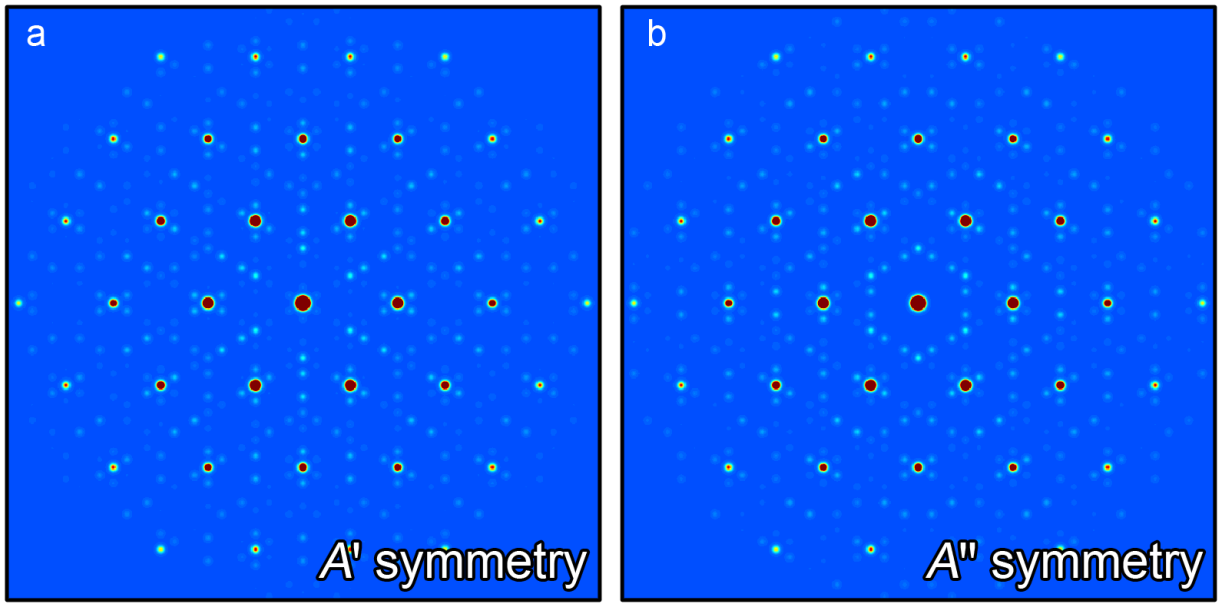

**Supplementary Figure 3 | Simulated Electron diffraction patterns of 1T-TaSeTe along the [001] zone axis. a** and **b** are simulated with incommensurate atomic displacements of $A'$ (eigenvector: [-0.0360, 0, 0; -0.0116$e^{1.095i}$, 0, 0; -0.0116$e^{-1.095i}$, 0, 0]) and $A''$ (eigenvector: [0, -0.0360, 0; 0, -0.0116$e^{1.095i}$, 0; 0, -0.0116$e^{-1.095i}$, 0]) symmetry, respectively. The simulations are based on the kinematic model. The intensities of CDW peaks around {110} spots shown in Fig. 2a are qualitatively consistent with pattern **a**, which indicates that the atomic displacements are predominantly longitudinal and have A' symmetry.

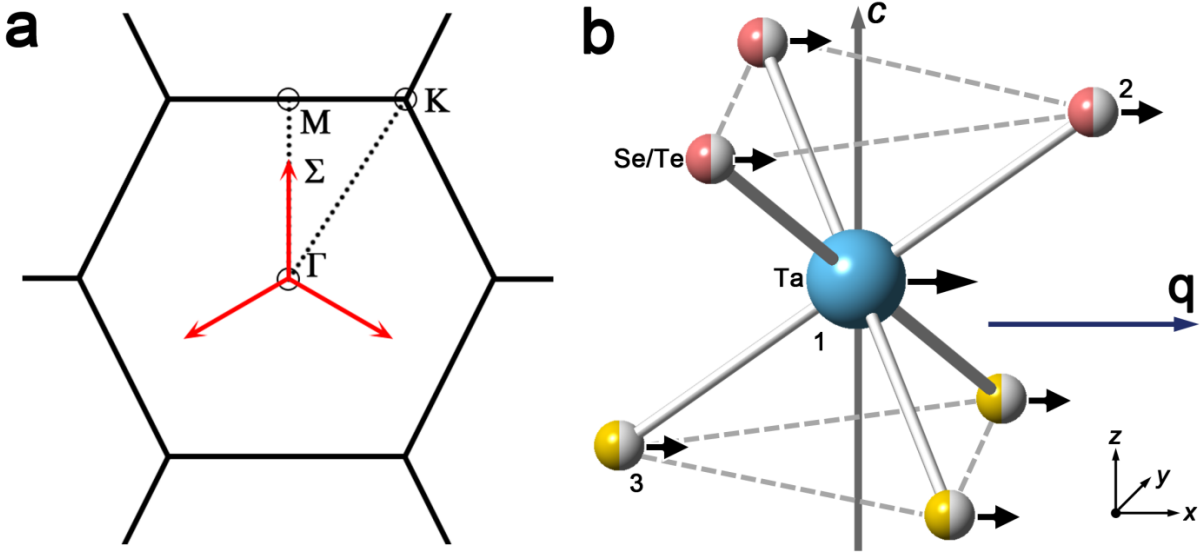

**Supplementary Figure 4 | In-plane Atomic displacements characteristic of A' symmetry. a,** Brillouin zone for 1T-TaSeTe in the normal phase (no lattice distortions). The red arrows indicate the in-plane wavevectors of three coexist CDWs. **b,** The number 1-3 correspond to a Ta atom in the middle, a Se/Te atom on the top plane and a Se/Te atom on the bottom plane, respectively. For A' symmetry, the atomic displacements within the Basal plane are all parallel to the **q** vector, indicating the longitudinal feature of this mode. The out-of-plane components are not shown in this figure because our experimental measurements can only reflect the in-plane atomic displacements. Supplementary Table 2 is constructed based on the atomic configuration shown here. According to the discussion in Section a, the amplitude of Ta displacement wave is $u_{Ta}$ and its phase is 0 at the origin; the amplitude and phase of the Se/Te displacement wave on the upper (lower) layer are $u_{Se/Te}$ and $\varphi_{Se/Te}$ ($u_{Se/Te}$ and $-\varphi_{Se/Te}$), respectively. It is worth noting that the real atomic displacements are the summation of three **q**-dependent components.

b) **Structure factor of incommensurate modulated crystal.** In the kinematic approximation, a structure factor $F_\mathbf{H}$ can be used to describe the amplitude and phase of electrons diffracted from the crystal plane **H**, even in incommensurate modulated structures[20,34]. In our case, the position of diffraction peaks in the reciprocal space are given by $\mathbf{H} = h_1\mathbf{a}^* + h_2\mathbf{b}^* + h_3\mathbf{c}^* + h_4\mathbf{q}_1 + h_5\mathbf{q}_2 + h_6\mathbf{q}_3$, and the corresponding structural factor can be written as

$$F_\mathbf{H} = \sum_{n,j} f_j \exp[2\pi i \mathbf{H} \cdot \boldsymbol{r}(\boldsymbol{n},j)]$$

$$= \sum_{n,j} f_j \exp[2\pi i \mathbf{H} \cdot (\boldsymbol{n} + \boldsymbol{r}_j)]$$

$$\times \exp\{2\pi i \mathbf{H} \cdot \mathbf{U}_{1j}\sin[2\pi \mathbf{q}_1 \cdot (\mathbf{n}+\mathbf{r}_j)+\varphi_{1j}]\}$$
$$\times \exp\{2\pi i \mathbf{H} \cdot \mathbf{U}_{2j}\sin[2\pi \mathbf{q}_2 \cdot (\mathbf{n}+\mathbf{r}_j)+\varphi_{2j}]\}$$
$$\times \exp\{2\pi i \mathbf{H} \cdot \mathbf{U}_{3j}\sin[2\pi \mathbf{q}_3 \cdot (\mathbf{n}+\mathbf{r}_j)+\varphi_{3j}]\}$$

(2)

where $f_j$ is the atomic scattering factor of atom $j$, $r(n, j)$ is the position of the $j$th atom in the modulated structure, its position in the normal state is given by the combination of the unit cell vector $\mathbf{n}$ and the atomic shift vector $\mathbf{r}_j$ within the unit cell, and the summation extends over all atoms in the illuminating area. $\mathbf{U}_{ij}$ is the displacement vector for the $j$th atom's modulation, as indicated by the arrows in Supplementary Fig. 4, and $\varphi_{ij}$ is the phase term as described in the previous section: $\varphi_{11}=\varphi_{21}=\varphi_{31}=0$, $\varphi_{12}=\varphi_{22}=\varphi_{32}=\varphi_{Se/Te}$ and $\varphi_{13}=\varphi_{23}=\varphi_{33}=-\varphi_{Se/Te}$. Using the Jacobi-Anger relation:

$$\exp(iz\sin\theta) \equiv \sum_{n=-\infty}^{\infty} J_n(z)\exp(in\theta)$$

(3)

where $J_n(z)$ is the $n$th-order Bessel function, equation (2) can be rewritten as

$$F_{\mathbf{H}} = \sum_{n,j} f_j \exp[2\pi i \mathbf{H} \cdot (\mathbf{n}+\mathbf{r}_j)]$$
$$\times \sum_{m_1=-\infty}^{\infty} J_{-m_1}(2\pi \mathbf{H} \cdot \mathbf{U}_{1j}) \cdot \exp\{-im_1[2\pi \mathbf{q}_1 \cdot (\mathbf{n}+\mathbf{r}_j)+\varphi_{1j}]\}$$
$$\times \sum_{m_2=-\infty}^{\infty} J_{-m_2}(2\pi \mathbf{H} \cdot \mathbf{U}_{2j}) \cdot \exp\{-im_2[2\pi \mathbf{q}_2 \cdot (\mathbf{n}+\mathbf{r}_j)+\varphi_{2j}]\}$$
$$\times \sum_{m_3=-\infty}^{\infty} J_{-m_3}(2\pi \mathbf{H} \cdot \mathbf{U}_{3j}) \cdot \exp\{-im_3[2\pi \mathbf{q}_3 \cdot (\mathbf{n}+\mathbf{r}_j)+\varphi_{3j}]\}$$
$$= \sum_{n,j}\sum_{m_1}\sum_{m_2}\sum_{m_3} f_j \exp[2\pi i(\mathbf{H}-m_1\mathbf{q}_1-m_2\mathbf{q}_2-m_3\mathbf{q}_3)\cdot(\mathbf{n}+\mathbf{r}_j)]$$
$$\times J_{-m_1}(2\pi \mathbf{H} \cdot \mathbf{U}_{1j}) \cdot J_{-m_2}(2\pi \mathbf{H} \cdot \mathbf{U}_{2j}) \cdot J_{-m_3}(2\pi \mathbf{H} \cdot \mathbf{U}_{3j})$$

$$\times \exp[-i(m_1\varphi_{1j} + m_2\varphi_{2j} + m_3\varphi_{3j})].$$

(4)

In addition, using the property

$$\sum_{n=-\infty}^{\infty} \exp[2\pi i s n] = \sum_{n=-\infty}^{\infty} \delta(s-n)$$

(5)

we can get

$$\begin{aligned}
\sum_{n} \exp[2\pi i \mathbf{S} \cdot \mathbf{n}] &= \sum_{n_1=-\infty}^{\infty} \sum_{n_2=-\infty}^{\infty} \sum_{n_3=-\infty}^{\infty} \exp[2\pi i (S_1 n_1 + S_2 n_2 + S_3 n_2)] \\
&= \sum_{n_1=-\infty}^{\infty} \sum_{n_2=-\infty}^{\infty} \sum_{n_3=-\infty}^{\infty} \delta(S_1-n_1) \cdot \delta(S_2-n_2) \cdot \delta(S_3-n_3) \\
&= \sum_{h=-\infty}^{\infty} \sum_{k=-\infty}^{\infty} \sum_{l=-\infty}^{\infty} \delta(S_1-h) \cdot \delta(S_2-k) \cdot \delta(S_3-l) \\
&= \sum_{\mathbf{G}} \delta(\mathbf{S}-\mathbf{G})
\end{aligned}$$

(6)

when the illuminated crystal can be considered to be infinitely large and the summation is over all Bragg peaks **G**. So the equation (4) can be simplified as

$$\begin{aligned}
F_{\mathbf{H}} &= \sum_{\mathbf{G},j} \sum_{m_1} \sum_{m_2} \sum_{m_3} f_j \delta(\mathbf{H} - m_1\mathbf{q}_1 - m_2\mathbf{q}_2 - m_3\mathbf{q}_3 - \mathbf{G}) \\
&\quad \times \exp[2\pi i (\mathbf{H} - m_1\mathbf{q}_1 - m_2\mathbf{q}_2 - m_3\mathbf{q}_3) \cdot \mathbf{r}_j] \\
&\quad \times J_{-m_1}(2\pi \mathbf{H} \cdot \mathbf{U}_{1j}) \cdot J_{-m_2}(2\pi \mathbf{H} \cdot \mathbf{U}_{2j}) \cdot J_{-m_3}(2\pi \mathbf{H} \cdot \mathbf{U}_{3j}) \\
&\quad \times \exp[-i(m_1\varphi_{1j} + m_2\varphi_{2j} + m_3\varphi_{3j})] \\
&= \sum_j f_j \exp[2\pi i \mathbf{K} \cdot \mathbf{r}_j] \cdot J_{-m_1}(2\pi \mathbf{H} \cdot \mathbf{U}_{1j}) \cdot J_{-m_2}(2\pi \mathbf{H} \cdot \mathbf{U}_{2j}) \cdot J_{-m_3}(2\pi \mathbf{H} \cdot \mathbf{U}_{3j})
\end{aligned}$$

$$\times \exp[-i(h_4\varphi_{1j} + h_5\varphi_{2j} + h_6\varphi_{3j})] \quad (7)$$

where **K** is the Bragg peak $\mathbf{K} = h_1\mathbf{a}^* + h_2\mathbf{b}^* + h_3\mathbf{c}^*$ associated with the scattering vector **H**.

The effects of atomic vibration can be characterized by the Debye-Waller factor which is dependent on the lattice temperature[22]. Considering the hexagonal-layered structure of 1T-TaSeTe, anisotropic Gaussian Debye-Waller factor is employed in our case

$$T_{\mathbf{H},j} = \exp[2\pi^2 \mathbf{H}^T \mathbf{U}_j \mathbf{H}] \quad (8)$$

where the symmetric atomic mean-square displacement tensor $\mathbf{U}_j$ is

$$\mathbf{U}_j = \begin{pmatrix} U_{11} & U_{12} & U_{13} \\ U_{12} & U_{22} & U_{23} \\ U_{13} & U_{23} & U_{33} \end{pmatrix}. \quad (9)$$

Because the effect of thermal fluctuation in the phase of the modulation wave (phasons) on the incommensurate diffraction satellites has been proved to be generally small, especially for the first-order satellites[35], this effect is ignored in our model.

**c) Intensity measurement and curve fitting.** Ten sets of diffraction patterns used for structural refinement were acquired at 26 K with pump fluence of 3.5 mJ/cm$^2$. The relative shifts (normally less than 2 pixels) among the patterns were measured and corrected based on cross-correlation analysis, and then the patterns with the same delay time were added up to improve the signal-noise-ratio. At each delay time, the intensities of 96 Bragg peaks and 84 satellite peaks (listed in Fig. 3a) were fitted by a two-dimensional rotated Gaussian function

$$FIT_\mathbf{H} = A\exp\left\{-\frac{[(x-x_0)\cos\theta - (y-y_0)\sin\theta]^2}{2\sigma_1^2} - \frac{[(x-x_0)\sin\theta + (y-y_0)\cos\theta]^2}{2\sigma_2^2}\right\}$$
$$+ (Bx + Cy + D) \quad (10)$$

where $(Bx+Cy+D)$ is used to estimate the local background intensity. In our experiments, the image size is 1024 × 1024 pixel, and we choose a 21 × 21 pixel block at each diffraction spot for the fitting procedure. The obtained value $I_\mathbf{H} = FIT_\mathbf{H} - (Bx + Cy + D)$ is regarded as the measured intensity of diffraction spot **H**. A typical fitting result is shown in Supplementary Fig. 5, indicating that equation (10) is a good approximation to depict the diffraction peak and background intensities (we also tried 2D Lorentzian and Moffat functions, but the Gaussian type gives the best fitting results in our case).

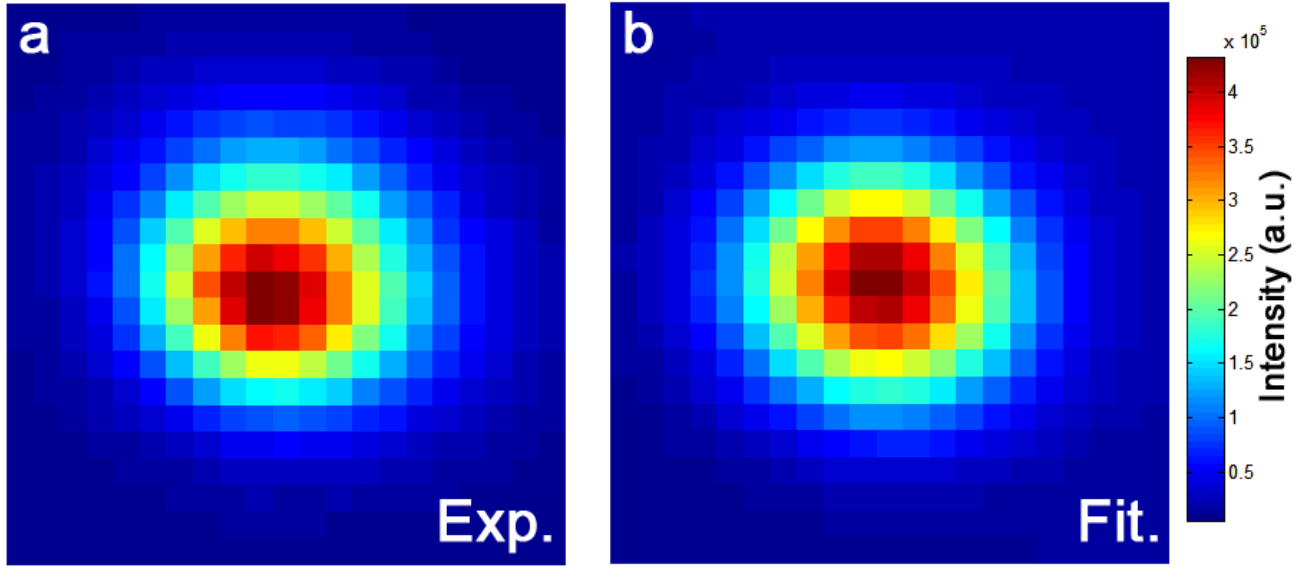

**Supplementary Figure 5 | Comparison between experimental intensity profile of (110) spot (a) and corresponding fitting result (b).** The fitting is based on 2D rotated Gaussian function.

The measured intensities of all Bragg peaks are summarized in Fig. 2b and c. In order to visually present the time evolutions of normalized intensities, we fitted the Type-1 and Type-2 data by two different double exponential functions

$$Func_1(t) = 1 + \frac{1}{2}\left[1 - \text{erf}\left(\frac{t_0 - t}{\tau_0}\right)\right] \cdot \left\{A \cdot \exp\left(\frac{-t}{\tau_1}\right) + B \cdot \left[1 - \exp\left(\frac{-t}{\tau_2}\right)\right]\right\} \quad (11)$$

$$Func_2(t) = 1 + \frac{1}{2}\left[1 - \text{erf}\left(\frac{-t}{\tau_0}\right)\right] \cdot \left\{A \cdot \left[1 - \exp\left(\frac{t_0 - t}{\tau_1}\right)\right]\right\}, \quad (12)$$

respectively, where erf(x) is the error function. The fitted curves are used as eye guides in this paper. The time dependent $u_{Ta}$ and $u_{Se/Te}$ are also fitted by equation (11), and $\varphi_{Se/Te}$ is fitted by linear function.

**d) Diffraction simulation and structural refinement.** For simulating electron diffraction patterns for quantitative analysis, both kinematic method and Bloch-wave method were used in our work. Under the assumptions of kinematical theory, the diffraction intensity of peak **H** can be given by $I_\mathbf{H} \propto |F_\mathbf{H}|^2$, where $F_\mathbf{H}$ is the structure factor calculated from equations (7) and (8). The atomic scattering factors of Ta, Se and Te were calculated based on the Gaussian fitting parameters provided in Ref. 36. In order to evaluate the effects of multiple scattering in diffraction patterns, the Bloch wave method[37,38] with incommensurate modulations and isotropic Debye-Waller factors was developed and used for simulations in Supplementary Fig. 2. In all simulated patterns, the 2D intensity profile of each diffraction peak was presented by a Moffat function

$$I_{peak} = I_\mathbf{H} \cdot \left[1 + \frac{(x-x_0)^2 + (y-y_0)^2}{\alpha^2}\right]^{-\beta} \tag{13}$$

where $(x_0, y_0)$ is the position of $\mathbf{H}$ and $\alpha = 10$ pixels, $\beta = 8$.

In order to extract the structure evolution after photoexcitation, the modulation parameters were refined by the nonlinear least-squares method which finds the minimum of $Q$ by the trust-region-reflective algorithm[39,40]

$$Q = \min\left\{\sum_\mathbf{H} [I_\mathbf{H}^{exp} - I_\mathbf{H}^{cal}(u_{Ta}, u_{Se/Te}, \varphi_{Se/Te}, U_{Ta}, U_{Se}, U_{Te})]^2\right\}$$

(14)

The summation is over all 180 measureable diffraction peaks whose weights are considered to be the same in our refinements. Both the experimental intensities $I_\mathbf{H}^{exp}$ and the fitted intensities $I_\mathbf{H}^{cal}$ are normalized by the most intense peaks {110}, i.e. $I_{\{110\}}^{exp} = I_{\{110\}}^{cal} \equiv 1$.

The quality of the fit to the diffraction data is evaluated by consideration of the value of normalized mean square error $R$

$$R = 1 - \left\|\frac{I_\mathbf{H}^{exp} - I_\mathbf{H}^{cal}}{I_\mathbf{H}^{exp} - mean(I_\mathbf{H}^{exp})}\right\|^2 \tag{15}$$

where || indicate the 2-norm of a vector and $R$ varies between $-\infty$ (bad fit) to 1 (perfect fit). A plot of $R$ vs. delayed time is shown in Supplementary Fig. 7, indicating that our structural model and refinement results are reasonably reliable.

We also tried to perform the refinement procedure based on the Bloch wave model in which the sample thickness/bending and the incident beam direction play important roles, making more independent parameters should be taken into consideration compared to the kinematic method and thus the refinements strongly dependent on the initial values chosen. On account of the significant bending/rippling of our sample within a large electron-illuminated area and the extremely high electron accelerating voltage, the kinematic theory is a good approximation to describe the diffraction intensities measured in UED, as evidenced by the comparison between two diffraction patterns acquired from UED and TEM shown in Supplementary Fig. 2. The time-consuming Bloch wave method thus has little advantage compared to the method used in our work.

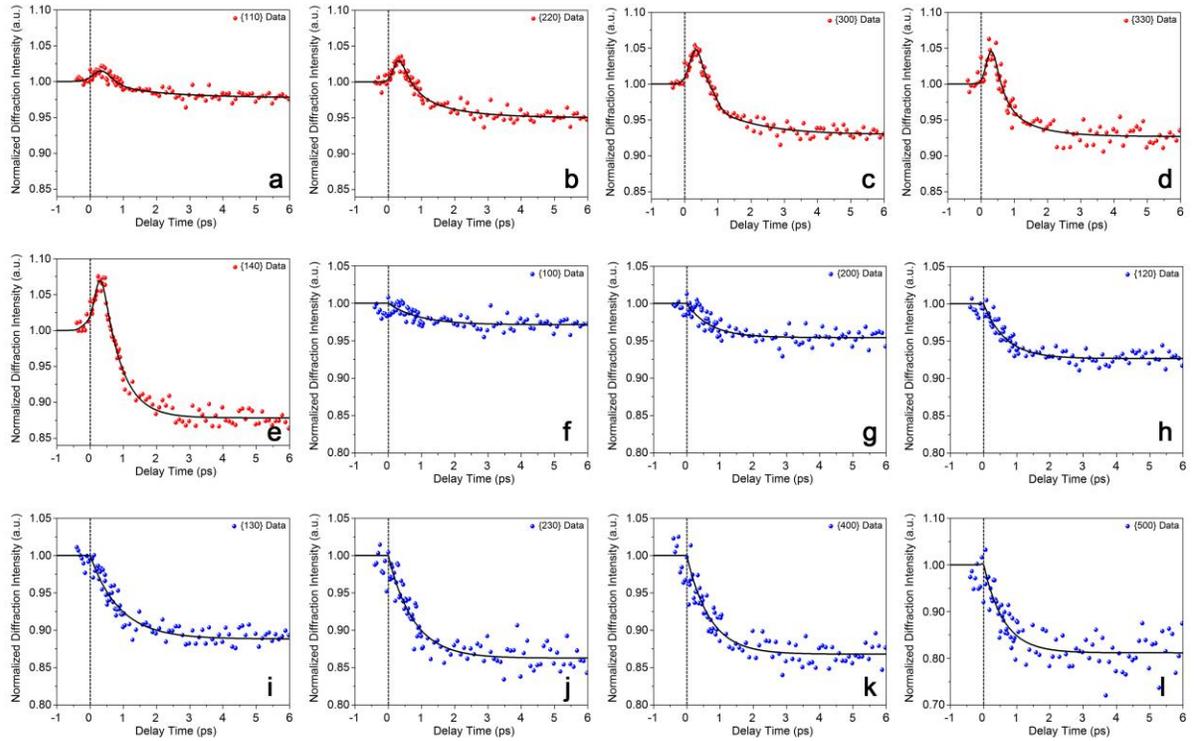

**Supplementary Figure 6 | Time dependence of Bragg peaks. a-e** Temporal evolution of the Type-1 Bragg peaks. **f-i** Temporal evolution of the Type-2 Bragg peaks. The pump was an 800 nm optical pulse at a fluence of 3.5 mJ/cm$^2$. The equations used for fitting the data are described in Section c.

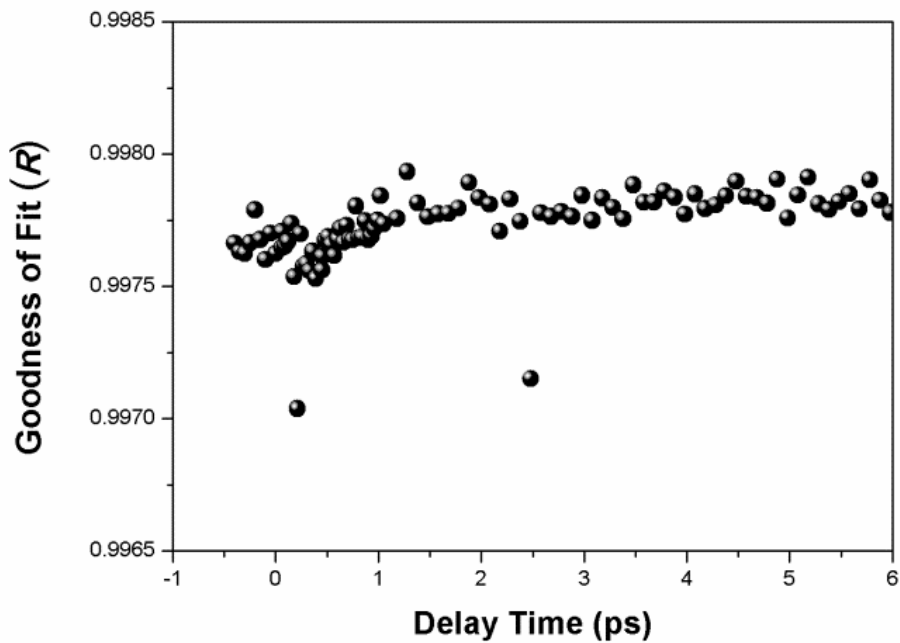

**Supplementary Figure 7 | Temporal evolution of *R* index which varies between –∞ (bad fit) to 1 (perfect fit).** The quality of the fit to the experimental data is evaluated by consideration of the *R* index defined in Section d. This plot indicates a good fit to the diffraction pattern at each delay time.

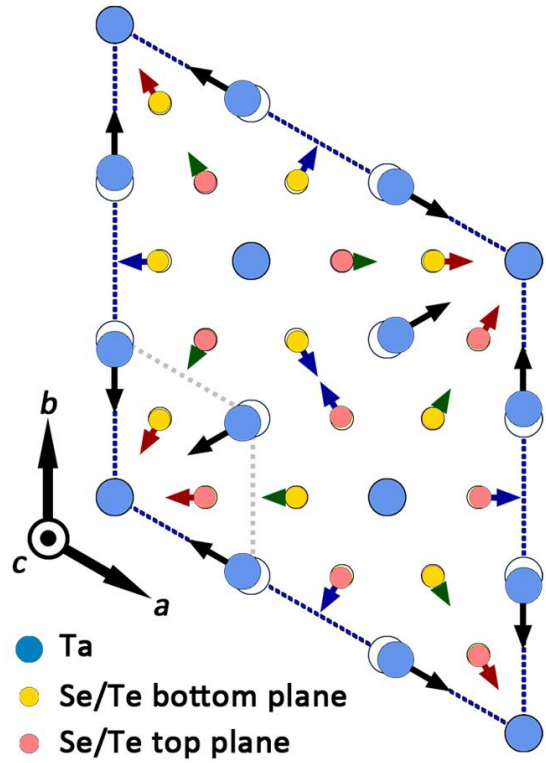

**Supplementary Figure 8 | DFT calculated basal plane projection of the displacements for a 3×3 simplified commensurate structure model of TaSeTe.** The atomic displacements correspond to the structural parameters $u_{Ta} = -0.04a$, $u_{Se/Te} = -0.005a$, $\varphi_{Se/Te} = 0.5$ rad and q = 1/3 (*a* is the length of normal state lattice vector ***a***). The atomic positions and unit cell in normal state are represented by the hollow circles and grey dashed lines. The arrows show the directions of atomic displacements and their lengths are exaggerated compared with the calculated values. Se/Te atoms in different layers are shown in two different colors. Considering the incommensurate feature in reality case, the structural parameters extracted from experimental (Fig. 3b) are slightly different from the calculated values. The calculation method is similar to the one used for 2H-TaSe$_2$ system described in Ref. 21.

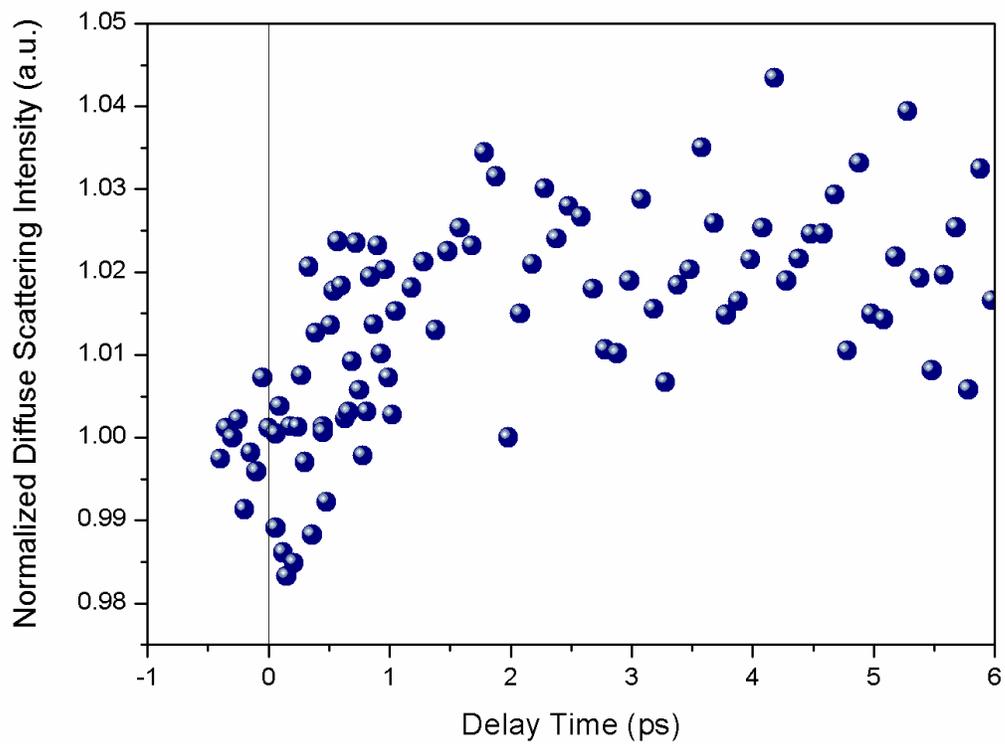

**Supplementary Figure 9 | Experimentally measured evolution of diffuse scattering (DS) intensity in UED patterns.** The increase of DS intensity indicates a gradual increasing of the lattice temperature.